\documentclass[prl,reprint,aps,superscriptaddress,nobibnotes,nofootinbib]{revtex4-1}

\usepackage[utf8]{inputenc}
\usepackage{mathptmx}
\usepackage{amsmath}
\usepackage{amssymb}
\usepackage[tight]{units}
\usepackage{color,graphicx}
\usepackage[colorlinks,citecolor=blue,urlcolor=blue,linkcolor=blue]{hyperref}
\usepackage[normalem]{ulem}
\usepackage{enumitem}
\usepackage{mathtools}
\usepackage{braket}

\newcommand{\cawo}{CaWO$_4$}
\newcommand{\ega}{$\beta$/$\gamma$}
\newcommand{\gev}{GeV/c$^2$}
\newcommand{\mev}{MeV/c$^2$}

\begin{document}
\title{First results from the CRESST-III low-mass dark matter program}
\newcommand{\mpi}{\affiliation{Max-Planck-Institut f\"ur Physik, 80805 M\"unchen, Germany}}
\newcommand{\coimbra}{\affiliation{Also at: LIBPhys, Departamento de Fisica, Universidade de Coimbra, P3004 516 Coimbra, Portugal}}
\newcommand{\hephy}{\affiliation{Institut f\"ur Hochenergiephysik der \"Osterreichischen Akademie der Wissenschaften, 1050 Wien, Austria}}
\newcommand{\ati}{\affiliation{Atominstitut, Technische Universit\"at Wien, 1020 Wien, Austria}}
\newcommand{\tum}{\affiliation{Physik-Department and Excellence Cluster Universe, Technische Universit\"at M\"unchen, 85747 Garching, Germany}}
\newcommand{\tuebingen}{\affiliation{Eberhard-Karls-Universit\"at T\"ubingen, 72076 T\"ubingen, Germany}} 
\newcommand{\oxford}{\affiliation{Department of Physics, University of Oxford, Oxford OX1 3RH, United Kingdom}}
\newcommand{\wmi}{\affiliation{Also at: Walther-Mei\ss ner-Institut f\"ur Tieftemperaturforschung, 85748 Garching, Germany}}
\newcommand{\lngs}{\affiliation{INFN, Laboratori Nazionali del Gran Sasso, 67010 Assergi, Italy}}
\newcommand{\gssi}{\affiliation{Also at: Gran Sasso Science Institute, 67100, L'Aquila, Italy}}
\newcommand{\cassino}{\affiliation{Also at: Dipartimento di Ingegneria Civile e Meccanica, Universit\'a degli Studi di Cassino e del Lazio Meridionale, 03043 Cassino, Italy}}
\newcommand{\chalmers}{\affiliation{Also at: Chalmers University of Technology, Department of Physics, 412 96 G\"oteborg, Sweden}}

\mpi
\lngs
\tum
\hephy
\ati
\tuebingen
\oxford

\coimbra
\gssi
\wmi
\cassino
\chalmers

\author{A.~H.~Abdelhameed}
  \mpi
  
\author{G.~Angloher}
  \mpi

\author{P.~Bauer}
  \email[Corresponding author: ]{philipp.bauer@mpp.mpg.de}
  \mpi

\author{A.~Bento}
  \mpi
  \coimbra 

\author{E.~Bertoldo}
  \mpi

\author{C.~Bucci}
  \lngs 

\author{L.~Canonica}
  \mpi 

\author{A.~D'Addabbo}
  \lngs
  \gssi

\author{X.~Defay}
  \tum 

\author{S.~Di~Lorenzo}
  \lngs
  \gssi

\author{A.~Erb}
  \tum
  \wmi
  
\author{F.~v.~Feilitzsch}
  \tum 
  
\author{S.~Fichtinger}
  \hephy

\author{N.~Ferreiro~Iachellini}
  \mpi  
  
\author{A.~Fuss}
  \hephy
  \ati

\author{P.~Gorla}
  \lngs 

\author{D.~Hauff}
  \mpi 

\author{J.~Jochum}
  \tuebingen 

\author{A.~Kinast}
  \tum
  
\author{H.~Kluck}
  \hephy
  \ati

\author{H.~Kraus}
  \oxford

\author{A.~Langenk\"amper}
  \tum

\author{M.~Mancuso}
  \mpi
  
\author{V.~Mokina}
  \hephy
  
\author{E.~Mondragon}
  \tum
  
\author{A.~M\"unster}
  \tum

\author{M.~Olmi}
  \lngs
  \gssi
  
\author{T.~Ortmann}
  \tum

\author{C.~Pagliarone}
  \lngs 
  \cassino

\author{L.~Pattavina}
  \tum
  \gssi

\author{F.~Petricca}
  \mpi 

\author{W.~Potzel}
  \tum 

\author{F.~Pr\"obst}
  \mpi

\author{F.~Reindl}
\email[Corresponding author: ]{florian.reindl@oeaw.ac.at}
  \hephy
  \ati

\author{J.~Rothe}
  \mpi 
  
\author{K.~Sch\"affner}
  \lngs 
  \gssi

\author{J.~Schieck}
  \hephy
  \ati 

\author{V.~Schipperges}
  \tuebingen
  
\author{D.~Schmiedmayer}
   \hephy
   \ati

\author{S.~Sch\"onert}
  \tum 
  
\author{C.~Schwertner}
  \hephy
  \ati

\author{M.~Stahlberg}
  \email[Corresponding author: ]{martin.stahlberg@oeaw.ac.at}
  \hephy
  \ati

\author{L.~Stodolsky}
  \mpi 

\author{C.~Strandhagen}
  \tuebingen

\author{R.~Strauss}
  \tum

\author{C.~T\"urko$\breve{\text{g}}$lu}
  \hephy
  \ati

\author{I.~Usherov}
  \tuebingen 

\author{M.~Willers}
  \tum 

\author{V.~Zema}
  \lngs
  \gssi
  \chalmers

\collaboration{CRESST Collaboration}
\noaffiliation

\begin{abstract}
  The CRESST experiment is a direct dark matter search which aims to measure interactions of potential dark matter particles in an earth-bound detector. With the current stage, CRESST-III, we focus on a low energy threshold for increased sensitivity towards light dark matter particles. In this manuscript we describe the analysis of one detector operated in the first run of CRESST-III (05/2016-02/2018) achieving a nuclear recoil threshold of \unit[30.1]{eV}. This result was obtained with a \unit[23.6]{g} \cawo~crystal operated as a cryogenic scintillating calorimeter in the CRESST setup at the Laboratori Nazionali del Gran Sasso (LNGS). Both the primary phonon/heat signal and the simultaneously emitted scintillation light, which is absorbed in a separate silicon-on-sapphire light absorber, are measured with highly sensitive  transition edge sensors operated at \mbox{\unit[$\sim15$]{mK}}. The unique combination of these sensors with the light element oxygen present in our target yields sensitivity to dark matter particle masses as low as \unit[160]{\mev}.  
\end{abstract}
\maketitle

\section{Introduction} \label{sec:introduction}

Today, the Standard Model of particle physics provides a widely consistent description of the visible matter in the Universe. However, the ever-growing precision of cosmological observations substantiates the finding that the visible matter contributes comparatively little to the matter density of the Universe which is, instead, dominated by dark matter. Numerous experiments strive to decipher the nature of dark matter, either by a potential production of dark matter particles in collisions of Standard Model particles, by searching for secondary Standard Model particles originating from the annihilation of dark matter particles, or by aiming at the direct observation of interactions of dark matter particles in earth-bound detectors. As of today, none of these three channels delivered an unambiguous hint for dark matter particles. 

Since, in particular, the mass of the dark matter particle(s) is a-priori unknown, direct searches for dark matter need to cover the widest possible mass range. This necessarily implies the use of different experimental techniques. In the standard scenario, assuming spin-independent and elastic scattering of dark matter particles off nuclei, liquid noble gas experiments take the lead in the high mass range. Solid-state or gas detectors are best suited for light (\unit[$\lesssim1$]{\gev}) dark matter due to their lower energy thresholds. For spin-dependent interactions superheated bubble chambers play an important role. 

The CRESST-III experiment operates scintillating \cawo~crystals as cryogenic calorimeters, simultaneously measuring a phonon/heat and a scintillation light signal. A distinctive feature of the phonon signal is a precise determination of the energy deposited in the crystal, independent from the type of particle interaction. This property, in combination with a low energy threshold, makes cryogenic calorimeters particularly suited for low-mass dark matter detection. Contrary to the phonon signal, the scintillation light strongly depends on the type of particle interaction, yielding event-by-event discrimination between the dominant background (\ega-interactions) and the sought-for nuclear recoils. Phonon and light signals are acquired by transition edge sensors (TESs) operated at around \unit[15]{mK} and read out by SQUID amplifiers~\cite{angloher_results_2012}. 

In this work we analyze data acquired with detector A, which has the  lowest threshold among all detectors operated in the first run of CRESST-III.

\section{CRESST-III Setup and Detector Design} \label{sec:cresstiiidetector}

\subsection{Experimental Setup}
CRESST is located in the Laboratori Nazionali del Gran Sasso (LNGS) underground laboratory in central Italy which provides an overburden against cosmic radiation with a water-equivalent of \unit[3600]{m} \cite{ambrosio_vertical_1995}. Remaining muons are tagged by an active muon veto with 98.7\%  geometrical coverage \cite{angloher2009commissioning}. In addition, the experimental volume is protected by concentric layers of shielding material comprising - from outside to inside - polyethylene, lead and copper. The polyethylene shields from environmental neutrons, while lead and copper suppress $\gamma$-rays. A second layer of polyethylene inside the copper shielding guards against neutrons produced in the lead or the copper shields.

A commercial $^3${He}/$^{4}$He-dilution refrigerator provides the base temperature of about \unit[5]{mK}. Cryogenic liquids (LN$_2$ and LHe) are refilled three times a week causing a down-time of about \unit[3]{h} per refill.

\subsection{CRESST-III Detector Design}


\begin{figure}[htb]
\includegraphics[width=\columnwidth]{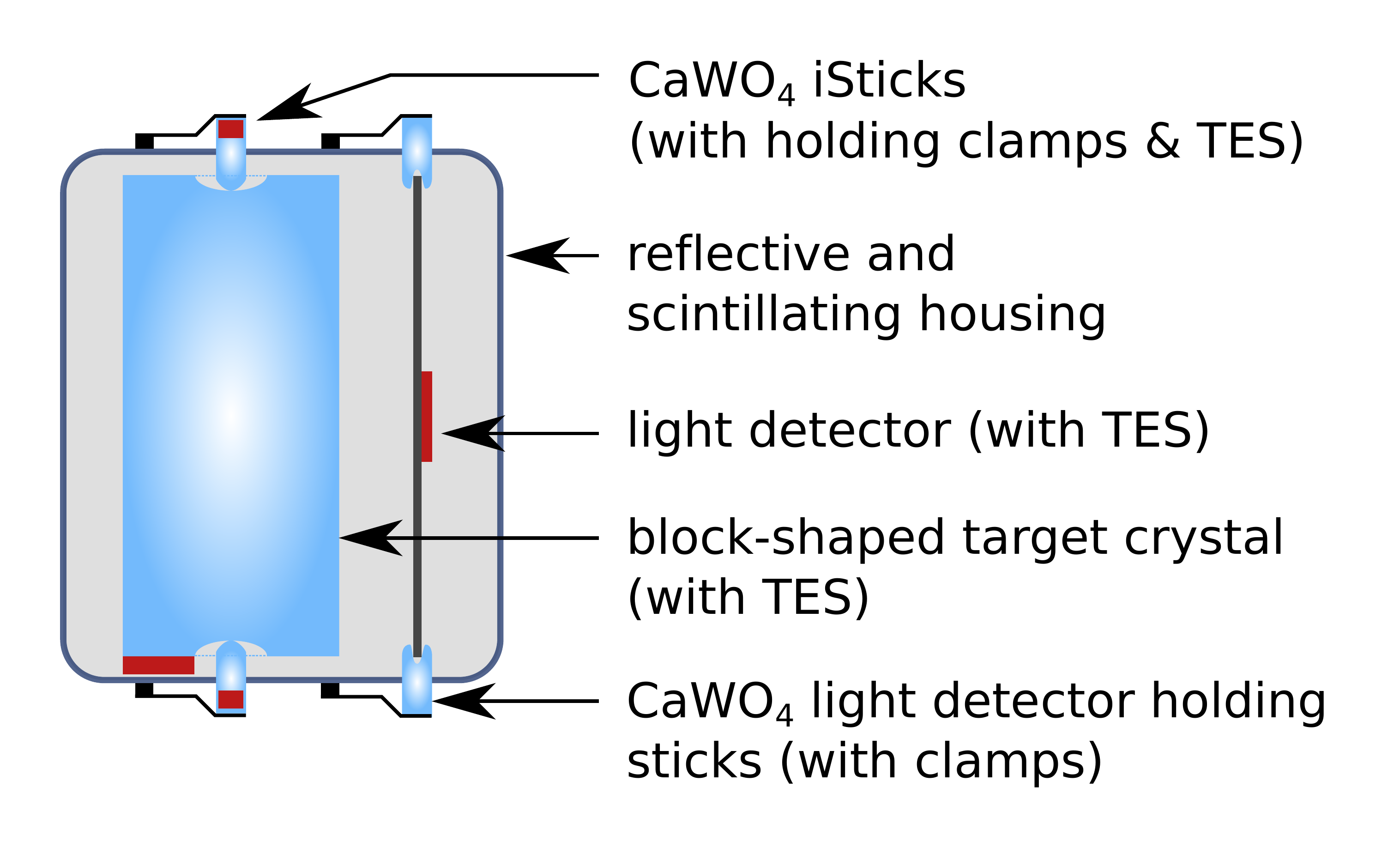}
\caption{Schematic of a CRESST-III detector module (not to scale). Parts in blue are made from \cawo, the TESs are sketched in red. The block-shaped target (absorber) crystal has a mass of \unit[$\sim$24]{g}, its dimensions are \unit[(20x20x10)]{mm$^{3}$}. It is held by three instrumented \cawo\ holding sticks (iSticks), two at the bottom and one on top. Three non-instrumented \cawo\ holding sticks  keep the square-shaped silicon-on-sapphire light detector in place. Its dimensions are \unit[(20x20x0.4)]{mm$^3$}.}
\label{fig:scheme}
\end{figure}

The \cawo~crystal of a CRESST-III detector module has a size of \unit[(20x20x10)]{mm$^3$} and a mass of \unit[$\sim$24]{g} (\unit[23.6]{g} for detector A). A schematic drawing is shown in figure \ref{fig:scheme}. The target crystal is held by three \cawo-sticks, each with a length of \unit[12]{mm}, a diameter of \unit[2.5]{mm} and a rounded tip of approximately \unit[2-3]{mm} in radius. The sticks are themselves instrumented with a TES, thus denoted iSticks. This novel, instrumented detector holder allows an identification and veto of interactions taking place in the sticks which might potentially cause a signal in the target crystal due to phonons propagating from the stick to the main absorber through their contact area. Since we veto interactions in any of the sticks, the three iSticks are connected in parallel to one SQUID, thus substantially reducing the number of necessary readout channels \cite{strauss_prototype_2017}.

Each target crystal is paired with a cryogenic light detector, matched to the size of the target crystal, consisting of a \unit[0.4]{mm} thick square silicon-on-sapphire wafer of \unit[20]{mm} edge length, also held by \cawo\ sticks and equipped with a TES. However, an instrumentation of these sticks is not needed as events within them would cause quasi light-only events which are outside the acceptance region for the dark matter search (see subsection \ref{subsec:eventselection}).\footnote{A small fraction of the light emitted by the stick might be absorbed by the target crystal creating a small phonon signal therein, thus these events are denoted quasi light-only.} 

The remaining ingredient to achieve a fully-active surrounding of the target crystal is the reflective and scintillating Vikuiti\texttrademark~foil encapsulating the ensemble of target crystal and light detector. Such a fully-active design ensures that surface-related backgrounds, in particular surface $\alpha$-decays, are identified and subsequently excluded from the dark matter analysis. A detailed study of the event classes arising from the iSticks and the light detector holding sticks is beyond the scope of this work; performance studies on the parallel TES readout may be found in \cite{rothe2016}.


%
\section{Dead-time free Recording and Offline Triggering} \label{sec:deadtimefreerecording}

In CRESST-III, the existing hardware-triggered data acquisition (DAQ) is extended by transient digitizers allowing for a dead-time free, continuous recording of the signals with a sampling rate of \unit[25]{kS/s}. Recording the full signal stream allows the use of an offline software trigger adapted to each detector. Our software trigger is based on the optimum filter or Gatti-Manfredi filter successfully used e.g.~by the CUORE experiment \cite{domizio_lowering_2011,alduino_low_2017}. The optimum filter maximizes the signal-to-noise ratio by comparing the frequency power spectrum of noise samples to that of an averaged pulse (a standard event). More weight is then given to pulse-like frequencies compared to those dominantly appearing in the noise samples. A full description of the method can be found in \cite{gatti1986processing}.

The complete stream is filtered with the optimum filter and a trigger is fired whenever the filter output for phonon or light channel exceeds a certain threshold value. For each channel we select a record window \unit[655.36]{ms} for further analysis. More details may be found in \cite{ferreiro_iachellini_increasing_2019}. The output of the optimum filter is not only used for the software triggering, but is also the basis of the energy reconstruction (see section \ref{sec:energyandevent}), yielding a precise value of the threshold in energy units.


\subsection{Optimal Trigger Threshold}
\label{subsec:triggerthreshold}
\begin{figure}[htb]
\includegraphics[width=\columnwidth]{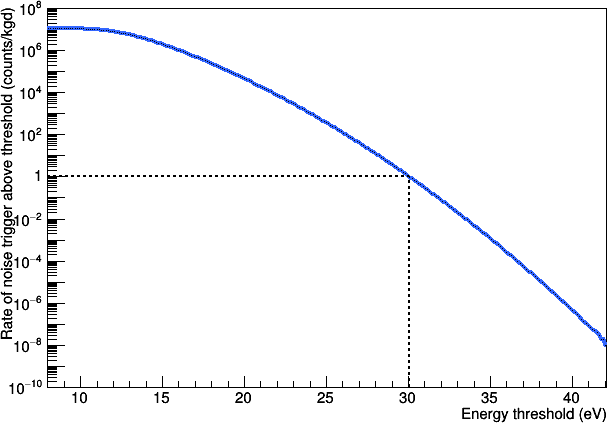}
\caption{Number of expected noise triggers surviving event selection per \unit[]{kg day} as function of a chosen trigger threshold for detector A. The threshold chosen for this work is indicated by the dashed line at \unit[30.1]{eV}.} 
\label{fig:triggercondition}
\end{figure}

Thanks to the continuously recorded data stream, the trigger threshold can be optimized based on a pre-defined criterion as described in \cite{mancuso_low_2018}, namely the number of noise triggers in a given time period (see figure \ref{fig:triggercondition}). For this analysis one noise trigger surviving event selection (see section \ref{subsec:eventselection}) per \unit[]{kg day} was allowed, corresponding to a trigger threshold of \unit[30.1]{eV}. 

\section{Energy Calibration and Event Selection} \label{sec:energyandevent}
\subsection{Energy Calibration} 
\label{subsec:energycalibration}
For CRESST-III, challenges arise from the greatly enhanced sensitivity. This leads to strong saturation effects at the \unit[122]{keV} $\gamma$-line from an external $^{57}$Co source used in former CRESST phases to calibrate the detectors. To directly probe the linear, non-saturated range of the detectors lower $\gamma$-ray energies would be required. Those, however, cannot efficiently penetrate the cryostat. Therefore, we perform an initial, approximate calibration using the K$_{\alpha 1}$ and K$_{\alpha 2}$ escape peaks of tungsten with a (weighted) mean energy of \unit[63.2]{keV} and later on fine-adjust by scaling to the \unit[11.27]{keV} peak (Hf L$_1$ shell, \cite{strauss_beta/gamma_2015}). The latter originates from cosmogenic activation of tungsten and is visible in all CRESST-III detectors (see figure \ref{fig:energyspectrum}).

The optimum filter offers a better resolution for the energy reconstruction than the standard event fit \cite{ferreiro_iachellini_increasing_2019}, as used in previous analyses. With the optimum filter we achieve a baseline resolution, i.e.~resolution at zero energy, of \unit[4.6]{eV}. However, saturation effects, that cannot be compensated by the optimum filter algorithm, set in at \unit[2.5]{keV} with complete signal saturation around \unit[75]{keV}. To partially overcome this limitation, we compare the amplitude determined by the optimum filter to the amplitude determined by a truncated standard event fit. The relation between these two quantities is obtained from the high statistics neutron calibration data and allows to extend the usable range for the optimum filter up to \unit[16]{keV}. Above \unit[16]{keV}, the saturation is too large to be reasonably corrected by this procedure and we therefore restrict our dark matter analysis to energies below \unit[16]{keV}.

\subsection{Light Yield Description using Neutron Calibration Data}

 To discriminate different types of particle interactions, we define the light yield of an event as the ratio of the energies deposited in light and phonon channel:
 $LY=E_l/E_p$. 

For this analysis, the phonon energy $E_p$ is considered to be the total deposited energy of an event; this approximation neglects any small possible dependence of $E_p$ on the event type and is motivated in appendix \ref{subsec:uncertainties}.
 
 Figure \ref{fig:neutroncalibration} shows the events surviving the selection criteria (see section \ref{subsec:eventselection}) in the AmBe neutron calibration data. The solid blue lines mark the \unit[90]{\%} upper and lower boundaries of the \ega-band. The red and green lines mark the bands expected for recoils off oxygen and tungsten, respectively. The calcium band lies in between the oxygen and the tungsten band and is not drawn for clarity. 

The description of the bands is done according to \cite{strauss_energy-dependent_2014}. The mean of the Gaussian \ega-band is given by a linear function plus a term accounting for the non-proportionality effect causing the bending down of the \ega-band towards low energies \cite{lang_scintillator_2009}. Quenching factors quantify the reduction in light output of a certain event type compared to a \ega-event of the same deposited energy. They were precisely measured in \cite{strauss_energy-dependent_2014} and allow to calculate the nuclear recoil bands. 

For the present work we fit the neutron calibration data (\ega-band plus nuclear recoil bands) utilizing an unbinned  maximum likelihood approach.  
Using the neutron calibration data, instead of the dark matter data directly, has several advantages. Firstly, it was found in \cite{strauss_energy-dependent_2014} that different crystals exhibit a slightly different quenching for nuclear recoils which, however, commonly affects all three nuclear recoil bands. In previous analyses we determined this common shift outside the likelihood fit by looking at oxygen scatters in neutron calibration data with energies above \unit[150]{keV}. The new likelihood fit, instead, directly obtains the common shift from the position of the nuclear recoil bands. The second advantage of performing the fit on neutron calibration data is a more populated \ega-band compared to dark matter data (compare figures \ref{fig:neutroncalibration} and \ref{fig:lightyield}). 

We would like to note that the term \ega-band should more correctly be denoted $\beta$-band, as $\gamma$-rays are known to produce slightly less scintillation light than $\beta$-particles~\cite{lang_scintillator_2009}. This is particularly apparent for the discrete $\gamma$-populations at \unit[2.6]{keV} and \unit[11.27]{keV} in figure \ref{fig:lightyield} that are clearly centered below the \ega-band. We model this effect in the new maximum likelihood fit. However, for means of clarity and convention, we stick to the term \ega-band.  

\begin{figure}[t]
  \includegraphics[width=\columnwidth]{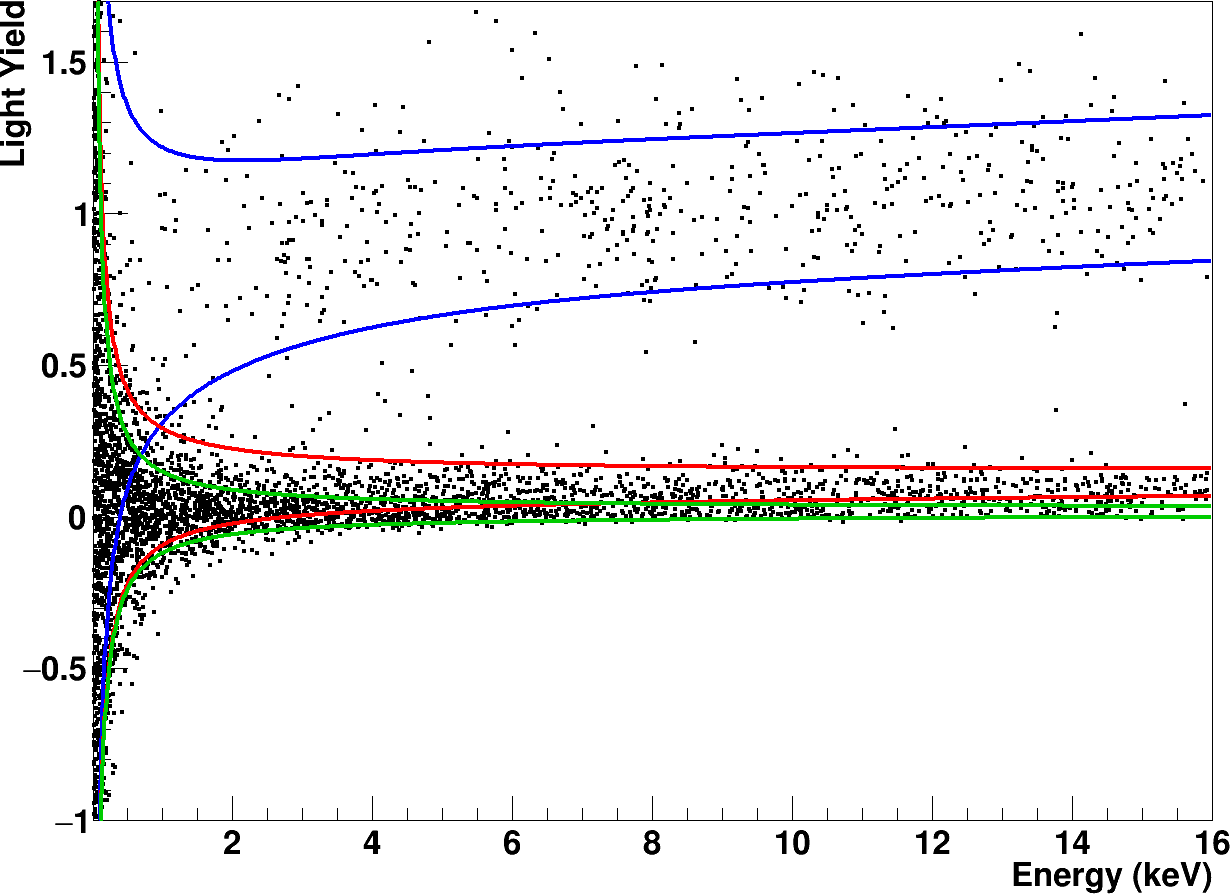}
  \caption{Neutron calibration data for detector A in the light yield versus energy plane. We fit these data to determine the bands for \ega-events (blue), nuclear recoils off oxygen (red) and tungsten (green), where the respective lines correspond to the upper and lower \unit[90]{\%} boundaries of the respective band. The band description follows \cite{strauss_energy-dependent_2014}.}
  \label{fig:neutroncalibration}
\end{figure}

The neutron calibration data also confirm that nuclear recoils and \ega-events have only a negligible pulse-shape difference. This justifies to use a single standard event as the basis for triggering and energy calibration.

\subsection{Data Pre-Selection} 

With stops in data taking for refills of cryogenic liquids, there are three data segments per week in standard data taking mode. In the rare cases of synchronization issues between the two data acquisitions (hardware-triggered and continuous) the affected segments are discarded. To establish the analysis procedure we define a non-blind training set by random selection of twenty percent of the collected data segments. This procedure is then blindly applied to the remaining \unit[80]{\%}. All results presented in this work, apart from the calibration steps, refer to the latter, ''blind'' dark matter dataset.

A binned rate cut is applied to remove periods of abnormally high rate, mainly due to small electronic disturbances, which were found to cluster in time. In total, this rate cut excludes \unit[14]{\%} of measuring time.

Through the injection of large heater pulses (control pulses) which heat the TES completely out of its transition, the current operating point in the transition curve is evaluated for each detector. The measured height of the control pulses is fed to an online feedback loop which maintains a constant operating point by adjusting the heating power accordingly. The control pulses are saved for offline analysis to remove time intervals where the measured control pulse height deviates by more than  \unit[3]{$\sigma$} from the mean of its Gaussian-shaped distribution. The additional amount of measuring time removed by this stability cut is \unit[3]{\%}.

\subsection{Event Selection} 
\label{subsec:eventselection}

Events where only the light channel triggered are removed from the dark matter analysis. The veto information from the instrumented \cawo\ sticks is exploited by removing all events with a pulse height in the iStick channel above noise. Additionally, we apply dedicated cuts to remove artifacts mimicking a pulse and events with distorted baselines which would potentially impair the energy reconstruction.

We quantify deviations of a real pulse from its nominal pulse shape during the application of the optimum filter algorithm. This is done by calculating the RMS difference between the filtered real pulse and the filtered standard event, where the latter is scaled to the amplitude of the real pulse. To increase the sensitivity to such deviations, we restrict this calculation to a window of \unit[$\pm30$]{ms} around the peak of the pulse. We call this quantity RMS\textsubscript{OF}. 

The RMS of the truncated standard event fit (RMS\textsubscript{SEF}) describes the agreement between a measured event and the standard event in the linear part of the detector response. This fit is performed simultaneously for phonon and light channel.

RMS\textsubscript{OF} was found to be less affected by changing noise conditions compared to RMS\textsubscript{SEF}, which indicates that RMS\textsubscript{OF} is more sensitive to the real pulse shape than RMS\textsubscript{SEF}. However, for energies above \unit[2.5]{keV} (see section \ref{subsec:energycalibration}) saturation effects start to corrupt RMS\textsubscript{OF} while RMS\textsubscript{SEF} remains unaffected. For this reason we apply cuts on both RMS quantities, aiming to remove events deviating from the nominal pulse shape.


A conservative muon veto cut is applied by rejecting detector events in a time interval of [\unit[-5]{ms}, \unit[+10]{ms}] around a muon veto trigger. Most of the events triggering the muon veto are not muons, but are due to radioactivity in the muon veto panels or the PMTs which does not penetrate the shielding. Thus, this cut almost exclusively removes randomly coincident  events. The total loss is \unit[7.6]{\%} of measurement time, which is consistent with the muon veto trigger rate of \unit[5.2]{Hz}. The cut values are conservatively based on considerations of detector time resolutions; no apparent correlation between muon panel hits and detector events was found.
An additional cut on coincidences between detector A and other detectors, using a coincidence window of \unit[$\pm10$]{ms}, is applied which causes negligible overall dead time and removes no further events in the acceptance region (see section \ref{sec:dmdata}). The main purpose of this cut is to remove potential neutron events which have a certain probability to cause energy deposits in multiple detectors, in contrast to dark matter particles.

The total exposure of the dark matter dataset after cuts amounts to \unit[3.64]{kg days}; the average survival probability for signal events, neglecting energy dependence, lies at approximately \unit[65]{\%} (see next section). 

\subsection{Efficiency / Signal Survival Probability} \label{subsec:efficiency}
\begin{figure}[h]
  \includegraphics[width=\columnwidth]{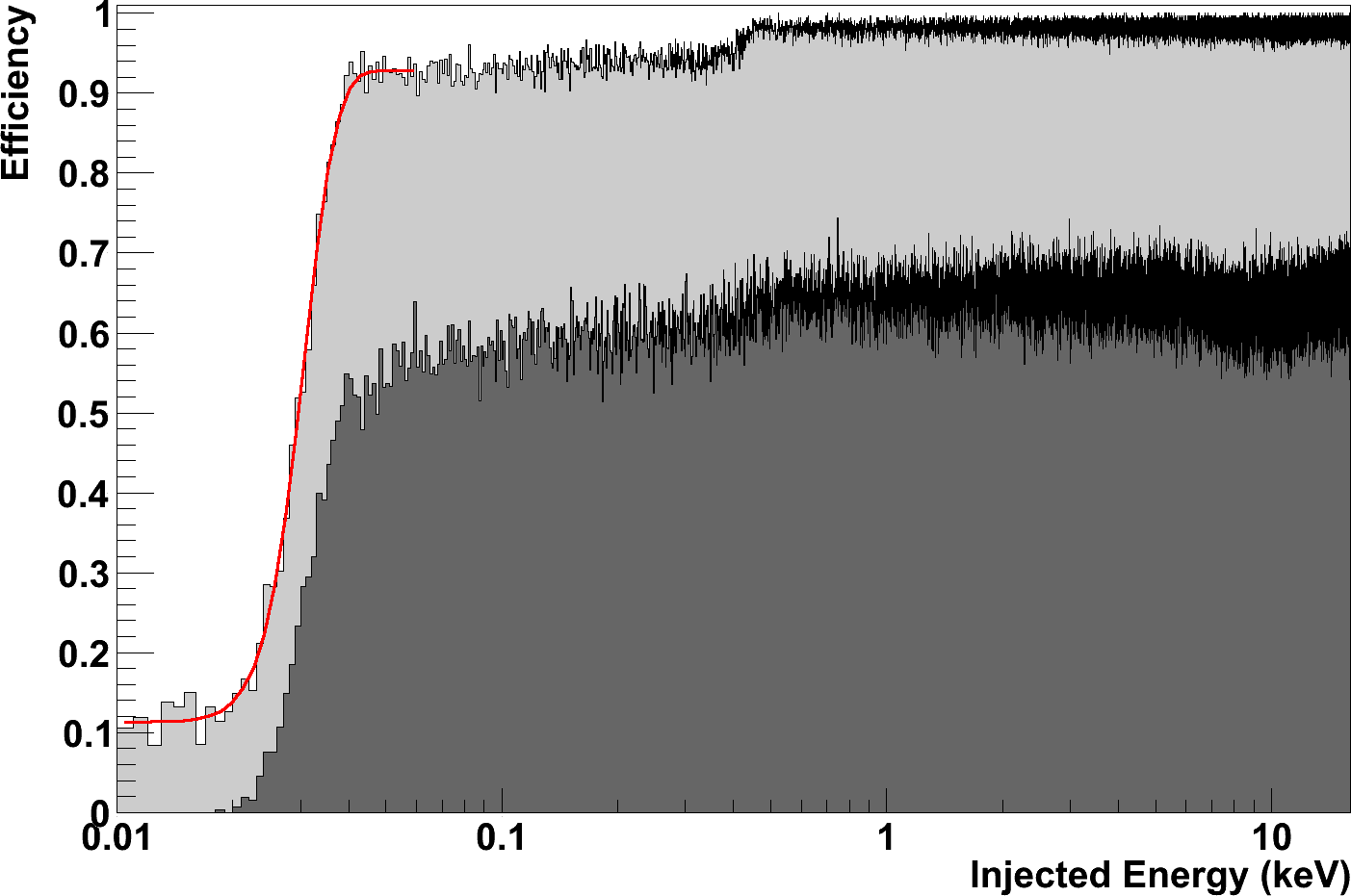}
  \caption{Efficiency obtained from simulated events and defined as probability for a valid signal event to be triggered (light gray) and pass the selection criteria (dark gray) as a function of injected (simulated) energy. The red line is a fit of the threshold with an error function, confirming the claimed value of \unit[30.1]{eV}.}
  \label{fig:efficiency}
\end{figure}

To determine the probability for a valid signal to be triggered and survive the selection criteria we pass simulated signal events through the complete analysis chain. The simulated events are created by superimposing the standard event onto the continuous data stream at randomly selected points in time. The events are simulated with high statistics ($\sim6.5\cdot10^{6}$ events for the full dark matter data set) and scaled in height corresponding to a flat energy spectrum from \unit[0]{} to \unit[20]{keV}.

As the timing of the simulated pulses is random and they are processed in the analysis chain exactly the same way as real pulses, the resulting loss in the simulated spectrum accounts for all artifacts on the stream, as well as cut effects, pile-up between events and dead-time due to injected heater pulses, providing an elegant and straightforward way of determining the signal survival probability. In addition, this procedure implicitly accounts for potential time-dependencies such as changing noise conditions. It should be noted, however, that pulse saturation effects are not taken into account in the simulation. This implies that the optimum filter amplitude for simulated pulses behaves strictly linearly, and a linearization using truncated fit results as discussed in section \ref{subsec:energycalibration} is not performed.

Figure \ref{fig:efficiency} shows the efficiency in bins of \unit[1]{eV} where the efficiency is defined as the ratio of surviving events to simulated events in the respective bin. The spectrum in light gray corresponds to all triggered events, the one in dark gray to events remaining after applying all selection criteria. As a cross-check we model the threshold with an error function depicted in red. Its fit yields a value for the threshold of \unit[($30.0\pm0.1$)]{eV} with a width of $\sigma=\unit[(5.3\pm0.2)]{eV}$; both values agree with expectations for the optimum trigger (see section \ref{subsec:triggerthreshold}) within uncertainties. Two features become apparent for the light gray trigger efficiency. Firstly, there is a pedestal of \unit[12]{\%} originating from pile-up of simulated events with previous large energy deposits or injected heater calibration pulses. Such events are efficiently rejected by our selection cuts, thus the pedestal vanishes for the efficiency after cuts. Secondly, pulses corresponding to energy deposits of less than \unit[$\sim0.4$]{keV} have a higher probability to be hidden by filter effects from a close-by optimal-filtered control pulse.\footnote{Typically, the optimum filter shows one global maximum at the position of the pulse and several local maxima before and after the main pulse \cite{domizio_lowering_2011,ferreiro_iachellini_increasing_2019}. The size of these local maxima for a control pulse (maximal possible pulse height) approximately equals the size of a \unit[0.4]{keV} energy deposition.}
  
In order to obtain a dark matter exclusion limit, we need to know what the expected dark matter signal looks like after triggering, energy reconstruction and event selection.  We simulate this by injecting artificial pulses into the continuous stream that follow the pulse height distribution of the expected recoil spectrum for each dark matter particle mass (the dark matter model will be discussed in section \ref{sec:results}). This method automatically includes all relevant aspects, in particular triggering efficiency and energy resolution, thus resulting in a dark matter recoil spectrum \textit{as it would be seen by our detector}. It should be explicitly noted that this newly implemented method overcomes the necessity of an analytic modeling of the detector response, in particular of the finite energy resolution. This represents a simplification in the extraction of dark matter results from the data, but above all avoids uncertainties introduced by the model of the detector response and/or the determination of the efficiencies of the analysis pipeline.  

To save computation time we perform only the aforementioned simulation, with a uniform energy distribution from \unit[0]{keV} - \unit[20]{keV}, and re-weigh each simulated event according to the expected recoil spectrum for a specific dark matter particle mass.

We ensure the robustness of our dark matter results by two conservative choices: Firstly, we do not make use of sub-threshold energies, i.e. energies below \unit[30.1]{eV}, where the trigger efficiency is \unit[50]{\%}.  Secondly, we reject simulated events whose reconstructed energies differ by more than two standard deviations from the injected/simulated energies. The latter criterion defends against the impact of single outliers caused by pile-up of a simulated event with a real particle event. Such pile-up may result in an overestimate of the survival probability of very small energy deposits.

\section{Dark Matter Dataset} \label{sec:dmdata}

The data used for dark matter analysis were taken between October 2016 and January 2018. The gross exposure before cuts is \unit[5.689]{kg days}.

\subsection{Light Yield} \label{subsec:lightyield}

\begin{figure}[t]
  \includegraphics[width=\columnwidth]{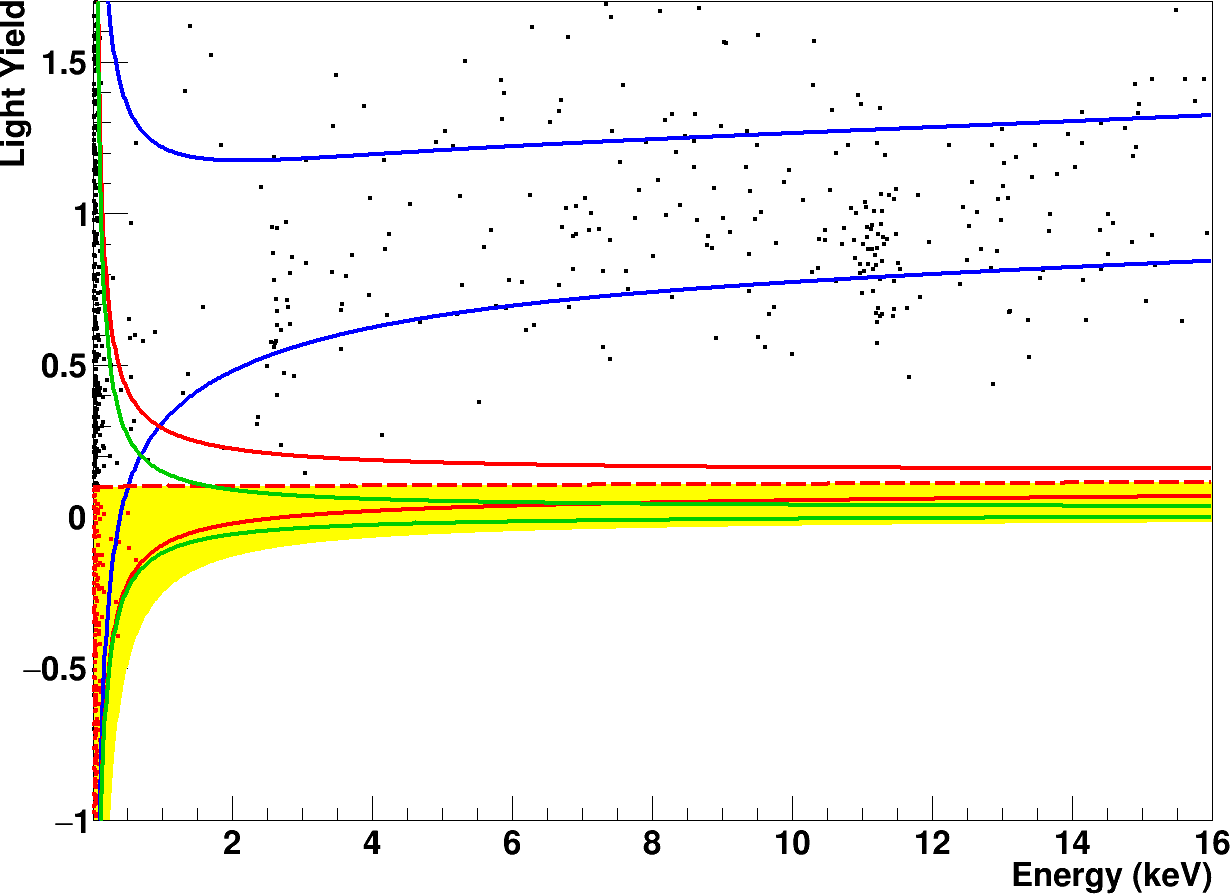}
  \caption{Light yield versus energy of events in the dark matter dataset, after selection criteria are applied (see section \ref{subsec:eventselection}). The blue band indicates the \unit[90]{\%} upper and lower boundaries of the \ega-band, red and green the same for oxygen and tungsten, respectively. The yellow area denotes the acceptance region reaching from the mean of the oxygen band (red dashed line) down to the \unit[99.5]{\%} lower boundary of the tungsten band. Events in the acceptance region are highlighted in red. The position of the bands is extracted from the neutron calibration data as shown in figure \ref{fig:neutroncalibration}.}
  \label{fig:lightyield}
\end{figure}

Figure \ref{fig:lightyield} shows the dark matter data after all the cuts described before in the light yield versus energy plane. In accordance with figure \ref{fig:neutroncalibration}, the blue, red and green bands correspond to \ega-events and nuclear recoils off oxygen and tungsten, respectively. The red dashed line depicts the mean of the oxygen band, which also marks the upper boundary of the acceptance region, shaded in yellow. The lower bound of the acceptance region is the \unit[99.5]{\%} lower boundary of the tungsten band, its energy span is from the threshold of \unit[30.1]{eV} to \unit[16.0]{keV}. Events in the acceptance region (highlighted in red) are treated as potential dark matter candidate events. We restrict the energy range to \unit[16]{keV} for this analysis since for higher energies the energy reconstruction cannot be based on the optimum filter method due to saturation effects. This choice, however, hardly affects the sensitivity for the low dark matter particle masses of interest.  The choice for the acceptance region was fixed a-priori before unblinding the data. We do not include the full oxygen recoil band in the acceptance region because the gain in expected signal is too small to compensate for the increased background leakage from the \ega-band. 

\begin{figure}[htbp]
  \includegraphics[width=\columnwidth]{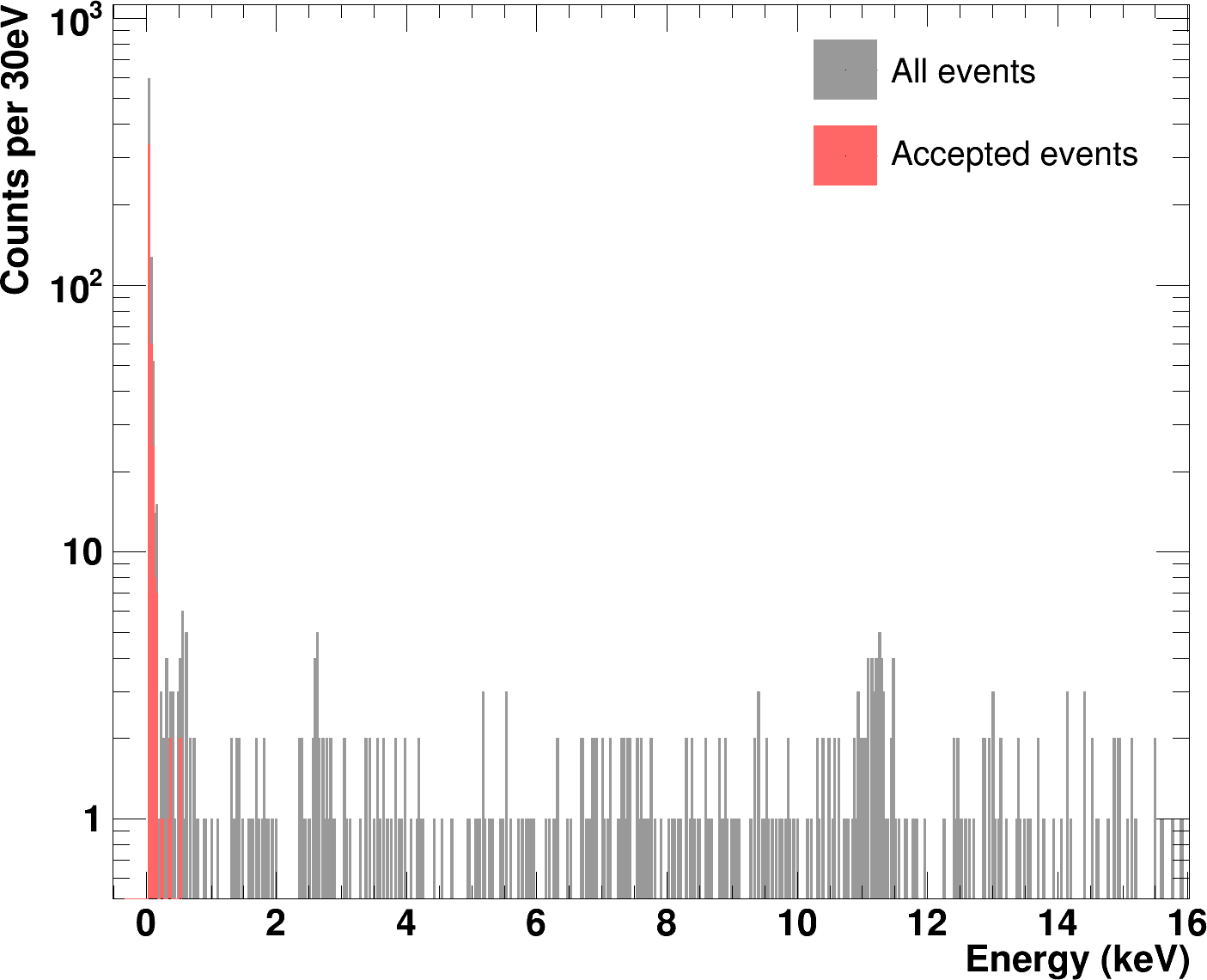}
  \caption{Energy spectrum of the dark matter dataset with lines visible at \unit[2.6]{keV} and \unit[11.27]{keV} originating from cosmogenic activation of $^{182}$W \cite{strauss_beta/gamma_2015}. Gray: all events, red: events in the acceptance region (see figure \ref{fig:lightyield}). }
  \label{fig:energyspectrum}
\end{figure}

\subsection{Energy Spectrum} 
\label{subsec:energyspectrum}

The corresponding energy spectrum is shown in figure \ref{fig:energyspectrum} with events in the acceptance region highlighted in red. In both figures \ref{fig:lightyield} and \ref{fig:energyspectrum}, event populations at \unit[2.6]{keV} and \unit[$\sim$11]{keV} are visible. These originate from cosmogenic activation of the detector material and subsequent electron capture decays:

\begin{equation*}
^{182}\text{W + p} \rightarrow{}^{179}\text{Ta} + \alpha, \ \   ^{179}\text{Ta} \xrightarrow[]{\text{EC}}{} ^{179}\text{Hf} + \gamma.
\end{equation*}

The latter decay has a half-life of 665 days, which implies a decreasing rate over the course of the measurement after initial exposure of the detector material. The energies of the lines correspond to the L\textsubscript{1} and M\textsubscript{1} shell binding energies of $^{179}$Hf with literature values of E\textsubscript{M$_1$}=\unit[2.60]{keV} and E\textsubscript{L$_1$}=\unit[11.27]{keV}, respectively \cite{firestone1999table}. As already mentioned in section \ref{subsec:energycalibration}, the clearly identifiable \unit[11.27]{keV} line was used to fine-adjust the energy scale, and therefore to give an accurate energy information in the relevant low-energy regime. These features were already observed in CRESST-II \cite{,angloher_results_2014, strauss_beta/gamma_2015}. Additionally, a population of events at \unit[$\sim$540]{eV} is visible, which hints at EC decays from the N\textsubscript{1} shell of $^{179}$Hf with a literature value of E\textsubscript{N$_1$}=\unit[538]{eV} \cite{firestone1999table}.

Below \unit[200]{eV}, an excess of events above the flat background is visible, which appears to be exponential in shape. Due to decreasing discrimination at low energies, it cannot be determined whether this rise is caused by nuclear recoils or \ega{} events (see figure \ref{fig:lightyield}). It should be emphasized that noise triggers are not an explanation for this excess, as it extends too far above the threshold of \unit[30.1]{eV}. According to the definition of the trigger condition in section \ref{subsec:triggerthreshold}, the expected number of noise triggers for the full dataset would be around 3.6. We observe an excess of events at lowest energies in all CRESST-III detector modules with thresholds below \unit[100]{eV}; the shape of this excess varies for different modules, which argues against a single common origin of this effect. No clustering in time of events from the excess populations is observed.

\begin{figure}[t]
  \includegraphics[width=\columnwidth]{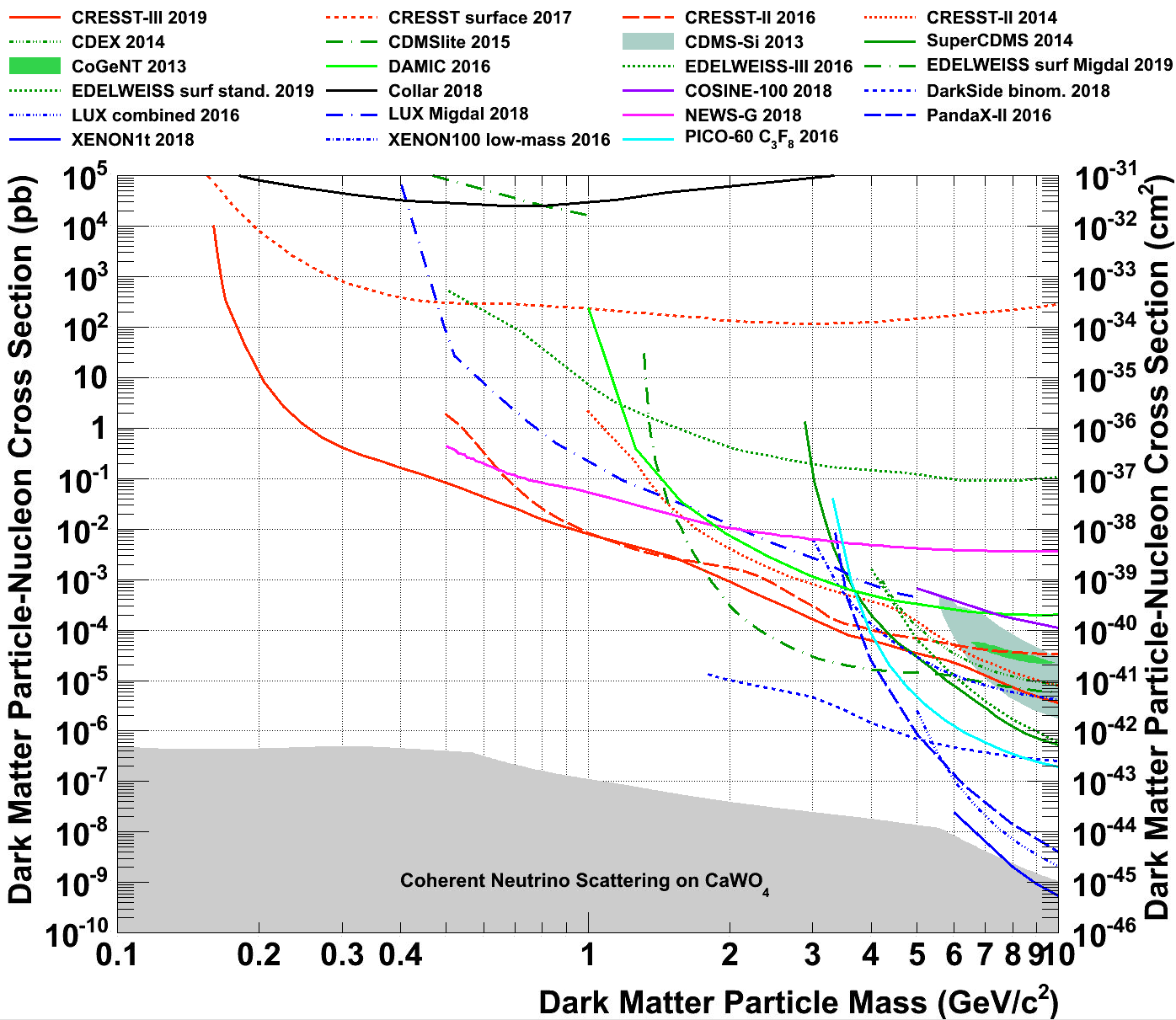}
  \caption{Experimental results on elastic, spin-independent dark matter nucleus scattering depicted in the cross-section versus dark matter particle mass plane. If not specified explicitly, results are reported with \unit[90]{\%} confidence level (C.L.). The result of this work is depicted in solid red with the most stringent limit between masses of \unit[(0.16-1.8)]{GeV/c$^2$}. The previous CRESST-II result is depicted in dashed red \cite{angloher_results_2016}, the red dotted line corresponds to a surface measurement performed with a gram-scale Al$_2$O$_3$ detector \cite{angloher_results_2017}.  We use a color-coding to group the experimental results: Green for exclusion limits (CDEX \cite{cdex_collaboration_limits_2014}, CDMSlite \cite{supercdms_collaboration_new_2016}, DAMIC \cite{damic_collaboration_search_2016}, EDELWEISS\cite{hehn_improved_2016,armengaud_searching_2019}, SuperCDMS \cite{supercdms_collaboration_search_2014}) and positive evidence (CDMS-Si (\unit[90]{\%}C.L.) \cite{supercdms_collaboration_search_2014}, CoGeNT (\unit[99]{\%}C.L.)\cite{aalseth_cogent:_2013}) obtained with solid state detectors based on silicon or germanium, blue for liquid noble gas experiments based on argon or xenon (DarkSide \cite{darkside_collaboration_low-mass_2018}, LUX \cite{lux_collaboration_results_2017,akerib_results_2018}, Panda-X\cite{pandax-ii_collaboration_dark_2017}, Xenon100\cite{xenon_collaboration_low-mass_2016}, Xenon1t\cite{xenon_collaboration_dark_2018}), violet for COSINE-100 (NaI) \cite{adhikari_experiment_2018}, black for Collar (H) \cite{collar_search_2018}, magenta for the gaseous spherical proportional counter NEWS-G (Ne + CH$_4$) \cite{arnaud_first_2018} and cyan for the super-heated bubble chamber experiment PICO (C$_3$F$_8$) \cite{pico_collaboration_dark_2016}. The gray region marks the so-called neutrino floor calculated for \cawo~in \cite{gutlein_impact_2015}. } 
  \label{fig:limit}


\end{figure}

\section{Results} \label{sec:results}

We use the Yellin optimum interval algorithm \cite{yellin_finding_2002,optimum_I} to extract an upper limit on the dark matter-nucleus scattering cross-section. In accordance with this method, we consider all 441 events inside the acceptance region to be potential dark matter interactions; no background subtraction is performed.

The anticipated dark matter spectrum follows the standard halo model \cite{donato_effects_1998} with a local dark matter density of ~{$\rho_\text{DM}$~=~\unit[0.3]{(GeV/c$^2$)/cm$^3$}, an asymptotic velocity of ~$v_\odot~=~\unit[220]{km/s}$ and an escape velocity of $v_\text{esc}~=~\unit[544]{km/s}$. Form factors, which are hardly relevant given the low transferred momenta here, follow the model of Helm~\cite{helm_inelastic_1956} in the parametrization of Lewin and Smith~\cite{lewin_review_1996}.

The result of the present analysis on elastic scattering of dark matter particles off nuclei is depicted in solid red in figure \ref{fig:limit} in comparison to the previous CRESST-II exclusion limit in dashed red and results from other experiments (see caption and legend of figure \ref{fig:limit} for details). The red dotted line corresponds to a surface measurement with a \unit[0.5]{g} Al$_2$O$_3$ crystal achieving a threshold of \unit[19.7]{eV} using CRESST technology~\cite{angloher_results_2017}.

The improvement in the achieved nuclear recoil threshold, in the respectively best performing detectors, from \unit[0.3]{keV} for CRESST-II to \unit[30.1]{eV} for CRESST-III, yields a factor of more than three in terms of reach for low masses, down to \unit[0.16]{GeV/c$^2$}. At \unit[0.5]{GeV/c$^2$} we improve existing limits by a factor of 6(30) compared to NEWS-G (CRESST-II). In the range  \unit[(0.5-1.8)]{GeV/c$^2$} we match or exceed the previously leading limit from CRESST-II.

\section{Conclusion}


In this paper, we report newly implemented data processing methods, featuring in particular the optimum filter technique for software-triggering and energy reconstruction. This allows one to make full use of the data down to threshold. The best detector operated in the first run of CRESST-III (05/2016-02/2018) achieves a threshold as low as \unit[30.1]{eV} and was, therefore, chosen for the analysis presented. 

In comparison to previous CRESST measurements, an indication of a $\gamma$-line at approximately \unit[540]{eV} compatible with the N$_1$ shell electron binding energy of $^{179}$Hf could be observed. Together with the reappearance of known lines, this corroborates the analysis of background components outlined in \cite{strauss_beta/gamma_2015}, as well as the energy calibration in this work. 

At energies below \unit[$200$]{eV} we observe a rising event rate which is incompatible with a flat background assumption and seems to point to a so-far unknown contribution. At the time of writing, dedicated hardware-tests with upgraded detector modules are underway to illuminate its origin.

We present exclusion limits on elastic dark matter particle-nucleus scattering, probing dark matter particle masses below \unit[0.5]{GeV/c$^2$} and down to \unit[0.16]{GeV/c$^2$}.

\section{Appendix}

\subsection{Study of Systematic Uncertainties} \label{subsec:uncertainties}

As discussed in section \ref{sec:energyandevent} the energy scale is adjusted using the \unit[11.27]{keV} $\gamma$-peak (Hf L$_1$ shell). As a consequence the energy scale is only strictly valid for events with a light yield of one. In particular, for a nuclear recoil less scintillation light is produced and, thus, more energy remains in the phonon channel leading to an overestimation of the phonon energy. Based on the fact that we measure both energies -- phonon ($E_p$) and light ($E_l$) -- one can account for this effect as was shown in \cite{angloher_results_2014} by applying the following correction:

 \begin{equation}
   E=\eta E_l + (1-\eta) E_p = [1-\eta (1-LY)]E_p.
   \label{eq:energyCorrection}
 \end{equation}
 In the above equation $\eta$ is the scintillation light efficiency which was determined to be \unit[$(6.6\pm0.4)$]{\%} in \cite{angloher_results_2014} for a crystal of the same origin as detector A, also grown within the CRESST collaboration.
 
However, for very low energies, the baseline noise of the light detector dominates the light signal, preventing a reasonable application of this correction. It should be noted that the resulting overestimation of the nuclear recoil energy scale is conservative as displayed in figure \ref{fig:uncertainty} where we show the exclusion limit after a reduction of \unit[7]{\%} of all event energies. This is the maximal possible systematic uncertainty introduced by not applying the correction outlined in equation \ref{eq:energyCorrection}. 

\begin{figure}[t]
  \includegraphics[width=\columnwidth]{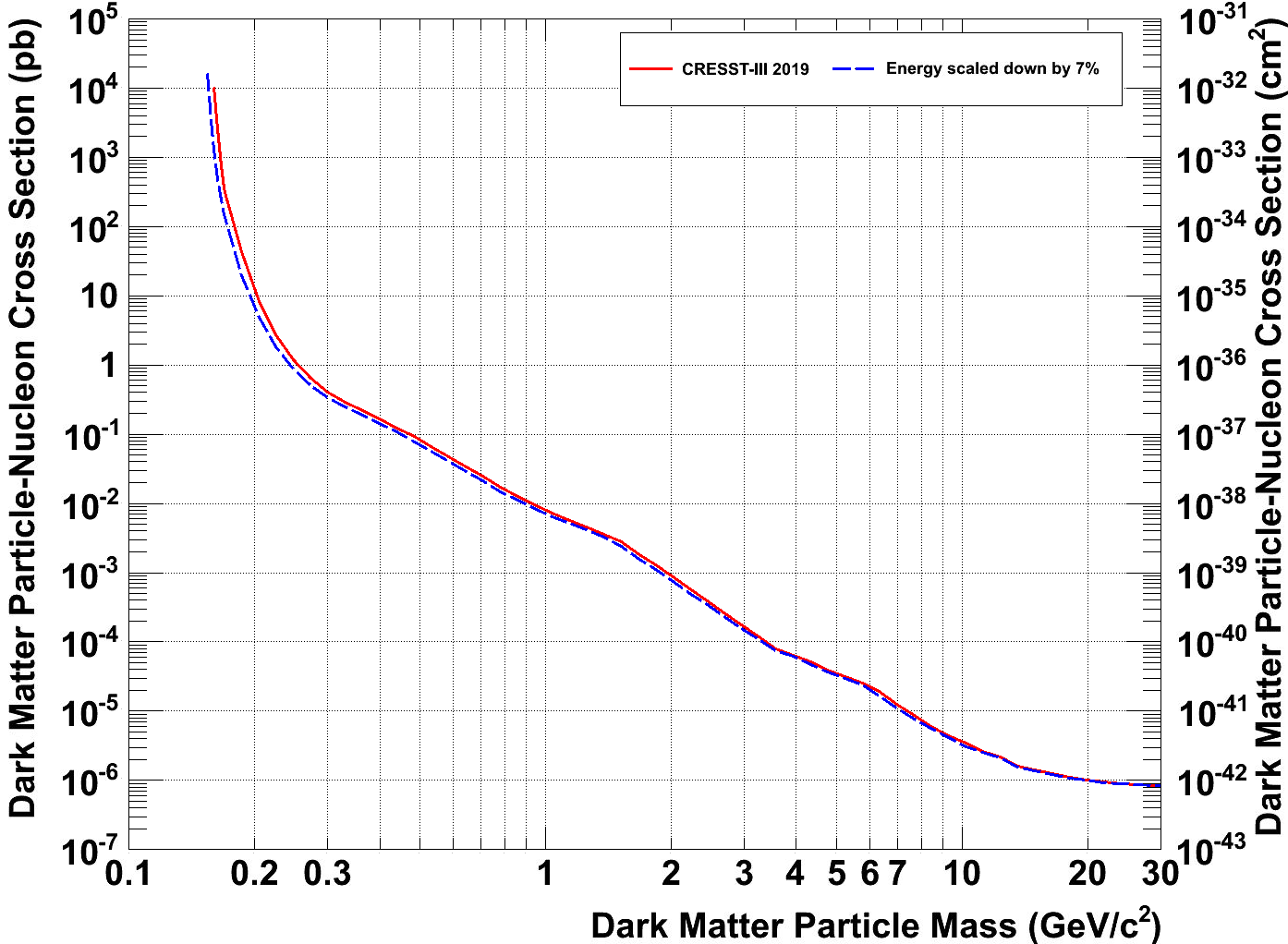}
  \caption{Illustration of the systematic uncertainty introduced by an overestimation of the nuclear recoil energy scale as a consequence of not correcting for the reduced scintillation light production of nuclear recoils via equation \ref{eq:energyCorrection}. Drawn are the result of this work in solid red (also see figure \ref{fig:limit}) and an exclusion limit obtained via scaling of the energy scale by \unit[7]{\%} (dashed blue line), which corresponds to the maximal possible overestimation. It can be seen that this systematic uncertainty has only a minor impact on the exclusion limit. Not applying the correction leads to a conservative exclusion limit for all dark matter particle masses.} 
  \label{fig:uncertainty}
\end{figure}

\subsection{Results on Spin-Dependent Interactions} \label{subsec:resultsSD}

In this article we present first results of CRESST-III on spin-independent elastic dark matter nucleus scattering. However, it deserves to be noted that the isotope $^{17}$O yields sensitivity for spin-dependent neutron-only interactions. The theoretical framework as well as the calculation of the expected rate exactly follows \cite{abdelhameed_first_2019} and, thus, just the result is given here. Compared to \cite{abdelhameed_first_2019}, values for the nuclear spin ($J=+5/2$), the mass ($A=17$) and the spin matrix element ($\braket{S_n} $=0.5)\cite{bednyakov_nuclear_2005,pacheco_nuclear_1989} are adjusted. We assume the $^{17}$O content to follow the minimal natural abundance of \unit[0.0367]{\%} \cite{holden_iupac_2018} which results in a gross $^{17}$O-exposure of only \unit[0.46]{g days}. Following \cite{hooper_robust_2018} and considering the rock composition of the LNGS overburden \cite{wulandari_neutron_2004}, we ensured that spin-dependent cross-sections of $\mathcal{O}$(\unit[10$^{9}$]{pb}) can be probed over the whole mass range under consideration. However, a more precise calculation of the upper boundary of the exclusion is subject to future work.

It should be stressed that nothing in the analysis chain but the signal expectation was changed when switching from the spin-independent to the spin-dependent case. The result is depicted in figure \ref{fig:SD17Limit} in solid red, together with a result from an above ground measurement of a Li$_2$MoO$_4$ crystal in dashed red and exclusion limits from CDMS-lite, LUX and Panda-X (see caption for references). 

Obviously, the small exposure for this measurement combined with the very low abundance of $^{17}$O results in a comparably modest limit for dark matter particles above \unit[1.5]{\gev}. However, the low nuclear recoil threshold of the presented detector A allows us to explore new parameter space for spin-dependent, neutron-only interactions from dark matter particle masses of \unit[1.5]{\gev} down to \unit[0.16]{\gev}.

\begin{figure}[t]
  \includegraphics[width=\columnwidth]{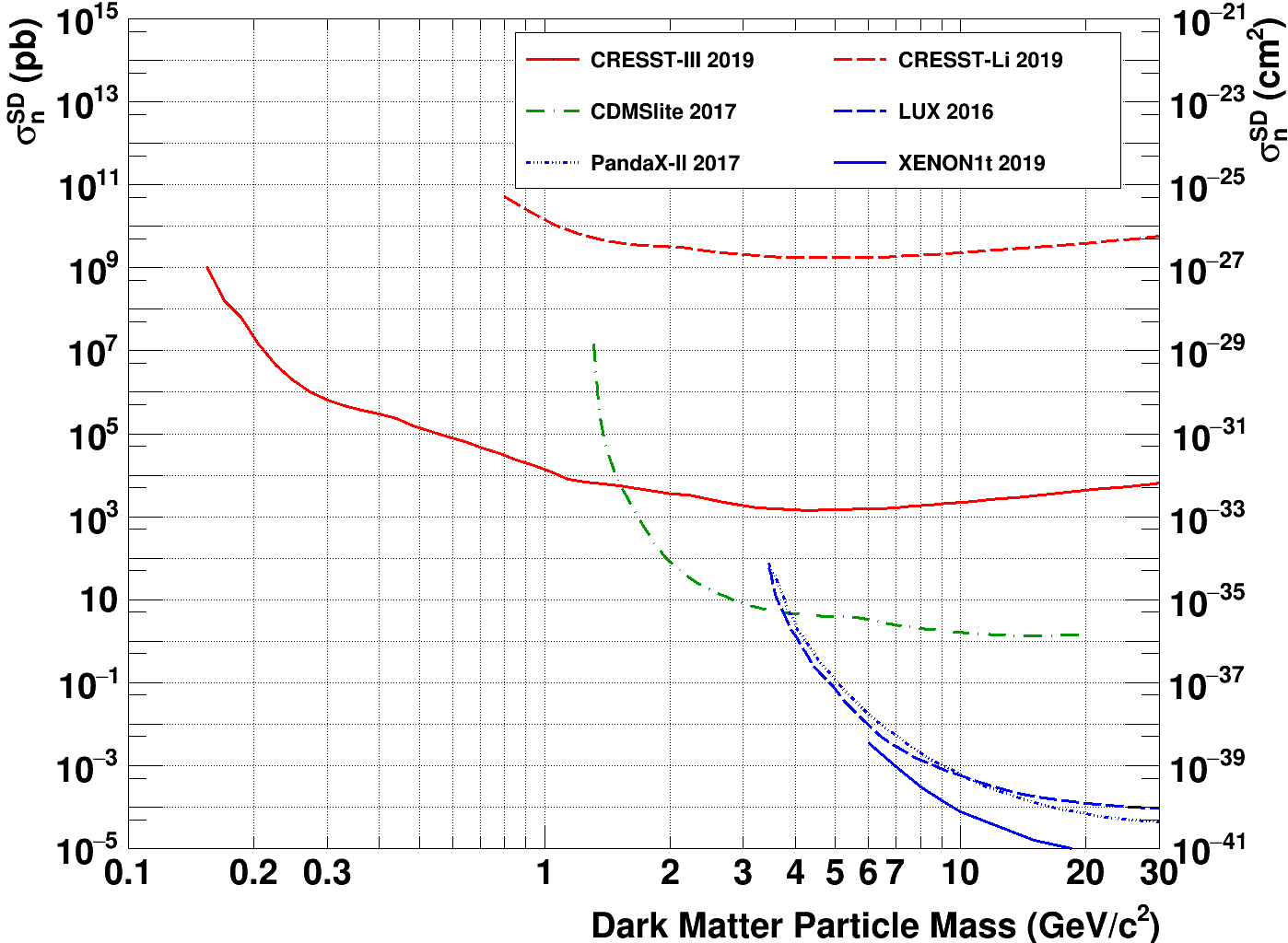}
  \caption{Results on spin-dependent neutron-only interactions via the isotope $^{17}$O in solid red (this work) and a result with $^7$Li in dashed red \cite{abdelhameed_first_2019}. Additionally, we plot results from CDMS-lite on $^{73}$Ge \cite{supercdms_collaboration_low-mass_2018}, LUX  \cite{lux_collaboration_results_2016}, Panda-X \cite{pandax-ii_collaboration_spin-dependent_2017} and XENON1t \cite{aprile_constraining_2019}, all three on $^{129}$Xe and $^{131}$Xe.}
  \label{fig:SD17Limit}
\end{figure}

\section*{Acknowledgements}
We are grateful to LNGS for their generous support of CRESST. We would like to thank Timon Emken and Riccardo Catena for the constructive discussion and their valuable input on the effect of earth-shielding.  This work has been supported through the DFG by the SFB1258 and the Excellence Cluster Universe, and by the BMBF: 05A17WO4 and 05A17VTA. 

We dedicate this work to our colleague Dr. Wolfgang Seidel who passed away suddenly and much too early in February 2017.

\bibliographystyle{h-physrev}
\bibliography{c3phase1.bib,run34paper.bib}
\end{document}